\documentstyle[twoside]{article}

\catcode`\@=11
\long\def\@makefntext#1{
\protect\noindent \hbox to 3.2pt {\hskip-.9pt  
$^{{\eightrm\@thefnmark}}$\hfil}#1\hfill}               %CAN BE USED 

\def\@makefnmark{\hbox to 0pt{$^{\@thefnmark}$\hss}}    %ORIGINAL 
\def\ps@myheadings{\let\@mkboth\@gobbletwo
\def\@oddhead{\hbox{}
\rightmark\hfil\eightrm\thepage}   
\def\@oddfoot{}\def\@evenhead{\eightrm\thepage\hfil
\leftmark\hbox{}}\def\@evenfoot{}
\def\sectionmark##1{}\def\subsectionmark##1{}}

\oddsidemargin=\evensidemargin
\newcounter{sectionc}\newcounter{subsectionc}\newcounter{subsubsectionc}
\renewcommand{\section}[1] {\vspace{12pt}\addtocounter{sectionc}{1} 
\setcounter{subsectionc}{0}\setcounter{subsubsectionc}{0}\noindent 
        {\tenbf\thesectionc. #1}\par\vspace{5pt}}
\renewcommand{\subsection}[1] {\vspace{12pt}\addtocounter{subsectionc}{1} 
        \setcounter{subsubsectionc}{0}\noindent 
        {\bf\thesectionc.\thesubsectionc. {\kern1pt \bfit #1}}\par\vspace{5pt}}
\renewcommand{\subsubsection}[1] {\vspace{12pt}\addtocounter{subsubsectionc}{1}
        \noindent{\tenrm\thesectionc.\thesubsectionc.\thesubsubsectionc.
        {\kern1pt \tenit #1}}\par\vspace{5pt}}
\newcommand{\nonumsection}[1] {\vspace{12pt}\noindent{\tenbf #1}
        \par\vspace{5pt}}

\newcounter{appendixc}
\newcounter{subappendixc}[appendixc]
\newcounter{subsubappendixc}[subappendixc]
\renewcommand{\thesubappendixc}{\Alph{appendixc}.\arabic{subappendixc}}
\renewcommand{\thesubsubappendixc}
        {\Alph{appendixc}.\arabic{subappendixc}.\arabic{subsubappendixc}}

\renewcommand{\appendix}[1] {\vspace{12pt}
        \refstepcounter{appendixc}
        \setcounter{figure}{0}
        \setcounter{table}{0}
        \setcounter{lemma}{0}
        \setcounter{theorem}{0}
        \setcounter{corollary}{0}
        \setcounter{definition}{0}
        \setcounter{equation}{0}
        \renewcommand{\thefigure}{\Alph{appendixc}.\arabic{figure}}
        \renewcommand{\thetable}{\Alph{appendixc}.\arabic{table}}
        \renewcommand{\theappendixc}{\Alph{appendixc}}
        \renewcommand{\thelemma}{\Alph{appendixc}.\arabic{lemma}}
        \renewcommand{\thetheorem}{\Alph{appendixc}.\arabic{theorem}}
        \renewcommand{\thedefinition}{\Alph{appendixc}.\arabic{definition}}
        \renewcommand{\thecorollary}{\Alph{appendixc}.\arabic{corollary}}
        \renewcommand{\theequation}{\Alph{appendixc}.\arabic{equation}}
        \noindent{\tenbf Appendix \theappendixc #1}\par\vspace{5pt}}
\newcommand{\subappendix}[1] {\vspace{12pt}
        \refstepcounter{subappendixc}
        \noindent{\bf Appendix \thesubappendixc. {\kern1pt \bfit #1}}
        \par\vspace{5pt}}
\newcommand{\subsubappendix}[1] {\vspace{12pt}
        \refstepcounter{subsubappendixc}
        \noindent{\rm Appendix \thesubsubappendixc. {\kern1pt \tenit #1}}
        \par\vspace{5pt}}

\newcommand{\textlineskip}{\baselineskip=13pt}
\newcommand{\smalllineskip}{\baselineskip=10pt}

\def\eightcirc{
\begin{picture}(0,0)
\put(4.4,1.8){\circle{6.5}}
\end{picture}}
\def\eightcopyright{\eightcirc\kern2.7pt\hbox{\eightrm c}} 

\newcommand{\copyrightheading}[1]
        {\vspace*{-2.5cm}\smalllineskip{\flushleft
        {\footnotesize International Journal of Modern Physics A, #1}\\
        {\footnotesize $\eightcopyright$\, World Scientific Publishing
         Company}\\
         }}

\newcommand{\publisher}[2]{{\begin{center}\footnotesize\smalllineskip 
        Received #1\\
        Revised #2
        \end{center}
        }}

\def\abstracts#1#2#3{{
        \centering{\begin{minipage}{4.5in}\baselineskip=10pt\footnotesize
        \parindent=0pt #1\par 
        \parindent=15pt #2\par
        \parindent=15pt #3
        \end{minipage}}\par}} 

\newcommand{\bibit}{\nineit}

\renewenvironment{thebibliography}[1]
        {\frenchspacing
         \ninerm\baselineskip=11pt
         \begin{list}{\arabic{enumi}.}
        {\usecounter{enumi}\setlength{\parsep}{0pt}
         \setlength{\leftmargin 12.7pt}{\rightmargin 0pt} %FOR 1--9 ITEMS
         \setlength{\itemsep}{0pt} \settowidth
        {\labelwidth}{#1.}\sloppy}}{\end{list}}

\newcounter{itemlistc}
\newcounter{romanlistc}
\newcounter{alphlistc}
\newcounter{arabiclistc}

\newcommand{\fcaption}[1]{
        \refstepcounter{figure}
        \setbox\@tempboxa = \hbox{\footnotesize Fig.~\thefigure. #1}
        \ifdim \wd\@tempboxa > 5in
           {\begin{center}
        \parbox{5in}{\footnotesize\smalllineskip Fig.~\thefigure. #1}
            \end{center}}
        \else
             {\begin{center}
             {\footnotesize Fig.~\thefigure. #1}
              \end{center}}
        \fi}

\newcommand{\tcaption}[1]{
        \refstepcounter{table}
        \setbox\@tempboxa = \hbox{\footnotesize Table~\thetable. #1}
        \ifdim \wd\@tempboxa > 5in
           {\begin{center}
        \parbox{5in}{\footnotesize\smalllineskip Table~\thetable. #1}
            \end{center}}
        \else
             {\begin{center}
             {\footnotesize Table~\thetable. #1}
              \end{center}}
        \fi}

\def\@citex[#1]#2{\if@filesw\immediate\write\@auxout
        {\string\citation{#2}}\fi
\def\@citea{}\@cite{\@for\@citeb:=#2\do
        {\@citea\def\@citea{,}\@ifundefined
        {b@\@citeb}{{\bf ?}\@warning
        {Citation `\@citeb' on page \thepage \space undefined}}
        {\csname b@\@citeb\endcsname}}}{#1}}

\newif\if@cghi
\def\cite{\@cghitrue\@ifnextchar [{\@tempswatrue
        \@citex}{\@tempswafalse\@citex[]}}
\def\citelow{\@cghifalse\@ifnextchar [{\@tempswatrue
        \@citex}{\@tempswafalse\@citex[]}}
\def\@cite#1#2{{$\null^{#1}$\if@tempswa\typeout
        {IJCGA warning: optional citation argument 
        ignored: `#2'} \fi}}

\def\pmb#1{\setbox0=\hbox{#1}
        \kern-.025em\copy0\kern-\wd0
        \kern.05em\copy0\kern-\wd0
        \kern-.025em\raise.0433em\box0}

\def\fnm#1{$^{\mbox{\scriptsize #1}}$}
\def\fnt#1#2{\footnotetext{\kern-.3em
        {$^{\mbox{\scriptsize #1}}$}{#2}}}

\def\fpage#1{\begingroup
\voffset=.3in
\thispagestyle{empty}\begin{table}[b]\centerline{\footnotesize #1}
        \end{table}\endgroup}

\def\runninghead#1#2{\pagestyle{myheadings}
\markboth{{\protect\footnotesize\it{\quad #1}}\hfill}
{\hfill{\protect\footnotesize\it{#2\quad}}}}
\font\tenrm=cmr10
\font\tenit=cmti10 
\font\tenbf=cmbx10
\font\bfit=cmbxti10 at 10pt
\font\ninerm=cmr9
\font\nineit=cmti9

\font\eightrm=cmr8

\def\qed{\hbox{${\vcenter{\vbox{                        %HOLLOW SQUARE
   \hrule height 0.4pt\hbox{\vrule width 0.4pt height 6pt
   \kern5pt\vrule width 0.4pt}\hrule height 0.4pt}}}$}}

\def\a{\alpha}

\def\s{\sigma}  
\def\d{\delta}

\def\m{\mu}  
\def\n{\nu}

\def\s9{{\scriptscriptstyle{9}}}
\def\h10{{\widehat{10}}}
\def\hA{{\hat A}}
\def\hG{{\hat G}}
\newcommand{\be}{\begin{equation}} 
\newcommand{\ee}{ \end{equation}}
\newcommand{\ba}{\begin{array}}
\newcommand{\ea}{\end{array}}
\newcommand{\bea}{\begin{eqnarray}}
\newcommand{\eea}{\end{eqnarray}}

\newcommand{\ft}[2]{{\textstyle\frac{#1}{#2}}}
\def\beqa{\begin{eqnarray}}
\def\eeqa{\end{eqnarray}}
\newcommand{\eqn}[1]{(\ref{#1})}

\begin{document}
%%%%%%%%%%%%%%%%%%%%%%%%%%%%%%%%%%%%%%%%%%%%%%%%%%%%%%%%%%%%%%%%%%%
\runninghead{M-Theory Duality and BPS-extended Supergravity} {M-Theory
Duality and BPS-extended Supergravity}  
%%%%%%%%%%%%%%%%%%%%%%%
\normalsize\textlineskip
\thispagestyle{empty}
\setcounter{page}{1}

\copyrightheading{}                     %{Vol. 0, No. 0 (1993) 000--000}

\vspace*{0.88truein}
%%%%%%%%%%%%%%%%%%%%%%%%%%%%%%%%%%%%%%
\fpage{1}
\centerline{\bf M-THEORY DUALITY AND BPS-EXTENDED SUPERGRAVITY}
%\vspace*{0.035truein}
%\centerline{\bf MANUSCRIPTS }
%\footnote{For
%the title, try not to use more than 3 lines. Typeset the title
%in 10 pt Times Roman, uppercase and boldface.}
%
\vspace*{0.37truein}
\centerline{\footnotesize BERNARD DE WIT}
%\footnote{Typeset names in
%10 pt Times Roman, uppercase. Use the footnote to indicate the
%present or permanent address of the author.}}
\vspace*{0.015truein}
\centerline{\footnotesize\it Institute for Theoretical Physics and
  Spinoza Institute}
\baselineskip=10pt
\centerline{\footnotesize\it Utrecht University, Utrecht, The
  Netherlands}
%\footnote{State completely without abbreviations, the
%affiliation and mailing address, including country. Typeset in 8
%pt Times Italic.}}
%\vspace*{10pt}
%\centerline{\footnotesize SECOND AUTHOR}
%\vspace*{0.015truein}
%\centerline{\footnotesize\it Group, Laboratory, Address}
%\baselineskip=10pt
%\centerline{\footnotesize\it City, State ZIP/Zone, Country}
\vspace*{0.225truein}
\publisher{(received date)}{(revised date)}
\vspace*{0.21truein}
\abstracts{We discuss toroidal compactifications of maximal
  supergravity coupled to an extended configuration of BPS
  states which transform consistently under the U-duality group. Under
  certain conditions this leads to theories that live in  
  more than eleven spacetime dimensions, with maximal supersymmetry 
  but only partial Lorentz invariance. We demonstrate certain features
  of this 
  construction for the case of nine-dimensional $N=2$ supergravity. 
  [ITP-UU-00/28; SPIN-00/26] }{}{}

%\textlineskip                  %) USE THIS MEASUREMENT WHEN THERE IS
%\vspace*{12pt}                 %) NO SECTION HEADING
%%%%%%%%%%%%%%%%%%%%%%%%%%%%%%%%%%%%%%%%%%%%%%%%%%%%%%%%%%
\vspace*{1pt}\textlineskip      %) USE THIS MEASUREMENT WHEN THERE IS
\section{Introduction}    %) A SECTION HEADING
\vspace*{-0.5pt}
\noindent
%%%%%%%%%%%%%%%%%%%%%%%%%%%%%%%%%%%%%%%%%%%%%%%%%%%%%%%%%%
Not too much is known about the underlying degrees of freedom of
M-theory. From perturbative string theory one knows about the existence
of massive and massless string states. The latter are captured in
supergravity where one can explore solitonic solutions which  
correspond to the branes that arise in (nonperturbative) string
theory. In spacetimes with compactified dimensions there exist winding 
and momentum states. These are BPS states, which can in principle be
incorporated into supergravity as matter
supermultiplets. As with all BPS states, one has a rather precise
knowledge about their mass spectrum and multiplicity, which suffices
for many applications. Massive string states decouple 
in the zero-slope limit; some of them are also BPS. More
daring ideas about the underlying degrees of freedom of M-theory are
based on supermembrane and matrix theories\cite{BST,DWHN,BFSS} (for
recent reviews, see refs.~4). 

M-theory is subject to a large number of duality
symmetries and equivalences. {From} the string perspective one knows
about S- and T-duality which are conjectured\cite{HullTownsend} to
be contained into the so-called U-duality group. This is an arithmetic
subgroup of the group of nonlinearly realized symmetries
(henceforth denoted by $\rm G$) of the 
maximal supergravity theories\cite{CremmerJulia}. Hence the
U-dualities coincide with the geometrical symmetries of 
supergravity. They are conjectured to be exact symmetries of
(toroidally compactified)  M-theory
and therefore act on the BPS states. For a comprehensive review,  see
ref.~7. The group G is expected to be broken down to
an integer-valued subgroup 
because it must respect the (perturbatively or nonperturbatively generated)
charge lattice of the BPS states. It is 
unavoidable that the inclusion of BPS states leads us back (at least 
in special limits) to eleven-dimensional supergravity\cite{CJS} or
extensions thereof, for the simple reason that the BPS states include
the Kaluza-Klein (KK) states needed to elevate the theory to eleven
dimensions. Of course, a multitude of arguments has already been
presented all pointing to the conclusion that
eleven-dimensional supergravity has a role to play (in particular, see
refs.~5,9,10).%{HullTownsend,Townsend,Witten}). 

Clearly, an important distinction between toroidally compactified
M-theory and toroidally compactified eleven-dimensional supergravity
resides in the BPS states.   
The compactifications of M-theory on a torus $T^n$ are conjectured to be
invariant under the duality group ${\rm E}_{n(n)}({\bf Z})$, which
includes the SL$(2,{\bf Z})$ S-duality group of the IIB superstring as
a subgroup. This conjecture includes the BPS states which must
transform according to the same duality group. In contrast with this, the
toroidal compactifications of eleven-dimensional 
supergravity generate towers of KK states whose central
charges are associated with momenta in the compactified 
dimensions. These states cannot transform under the duality
symmetry, simply because the central charges that they carry are
too restricted; central charges associated with the two- and
five-brane charges in eleven dimensions vanish for the KK
states. Obviously the massless states coincide for the two theories.  
But in view of the fact that M-theory contains more BPS states than
the KK states associated with the momenta on the hyper-torus, one may
wonder why the spacetime dimension should 
remain restricted to eleven. This is one of the questions that we will
address here. 

In principle, the central charges transform according to the automorphism 
group ${\rm H}_{\rm R}$ of the supersymmetry algebra that rotates the
supercharges and commutes with the $D$-dimensional Lorentz group,
where $D=11-n$. In most cases, however, 
the central charges can also be assigned to the larger U-duality group.
However, there are exceptions for spacetime dimensions
$D=11-n\leq5$ for the low-dimensional branes (for a discussion, see
refs.~7,11).% \cite{ObersPioline,istanbul}). 
On the other hand we note that the
antisymmetric gauge fields of supergravity are assigned to
representations of G and 
not of  ${\rm H}_{\rm R}$ (the only exceptions exist in $D=4,8$ where
one has to rely on the Hodge duality rotations of the field strengths
in order to realize the full duality group).   

When the central charges, and thus the corresponding BPS states that
carry those charges, constitute representations of the U-duality
group, it may be possible to incorporate all these BPS states into a
local supergravity field theory in a way that is U-duality
invariant. We call such theories BPS-extended  
supergravities. The construction of these theories is
motivatied in part by some old intriguing result that the toroidal 
compactifications of eleven-dimensional supergravity, without
truncation to the massless modes, exhibit traces of
the hidden symmetries\cite{deWitNicolai,Nicolai}, while one of the
reasons for the
lack of  invariance of the untruncated theory is precisely the 
incompleteness of the massive KK states with respect to the
hidden symmetries. Interestingly enough, by insisting on U-duality for
the BPS states 
one usually goes beyond eleven-dimensional supergravity.  To study the
properties  of these new theories we discuss the simplest version
corresponding to a toroidal compactification to nine spacetime
dimensions,  where we couple a variety of BPS supermultiplets to the
unique $N=2$ supergravity theory, which is thus based on 32
supercharges. The supergravity theory has been studied
previously\cite{BHO}, but the main emphasis here is on the BPS
states. We  also 
make contact with earlier work\cite{JHS} where it was argued that the
BPS multiplets have a natural interpretation in terms of the momentum
and wrapping states on the M-theory torus. The IIA/B T-duality and
the IIA/M S-duality combines as a duality between IIB theory on
$R^9\times S^1$ and M-theory on $R^9\times T^2$. As we were alluding
to above, this theory can be interpreted as a theory living in twelve
dimensions (although there is no twelve-dimensional Lorentz
invariance). In special decompactification limits it coincides with 
eleven-dimensional or IIA/B  supergravity\cite{ADLN}. In the next
sections we discuss various aspects of this nine-dimensional
BPS-extended theory. 
%%%%%%%%%%
%\textheight=7.8truein
%\setcounter{footnote}{0}
%\renewcommand{\thefootnote}{\alph{footnote}}

%%%%%%%%%%%%%%%%%%%%%%%%%%%%%%%%%%%%%%%%%%%%%%%%%%%%%%
\section{Maximal Supersymmetry and BPS States in Nine Dimensions}
\noindent
Let us start by considering the BPS multiplets that are relevant in nine
spacetime dimensions from the perspective of supergravity, string
theory and (super)membranes. It is well known that the massive
supermultiplets of IIA and IIB string theory 
coincide, whereas the massless states comprise inequivalent
supermultiplets, for the simple reason that they transform according
to different representations of the SO(8) helicity group.
When compactifying the theory on a circle, massless IIA and
IIB states in nine
spacetime dimensions transform according to identical SO(7)
representations of the helicity group and constitute equivalent
supermultiplets. The 
corresponding interacting field theory is the 
unique $N=2$ supergravity theory in nine spacetime dimensions. 
However, the supermultiplets of the BPS states, which carry momentum
along the circle, remain inequivalent, as they remain assigned to
the inequivalent representations of the group SO(8) which is now
associated with the restframe (spin) rotations of the massive states. 
The momentum states of the IIA and the IIB theories will be denoted
henceforth as KKA and KKB states, respectively. The fact that they
constitute  inequivalent supermultiplets, has implications for the
winding states in order that T-duality will remain valid. This is
discussed below. 

In ref.~16 %\cite{ADLN} 
this question was investigated in detail and 
considered in the context of $N=2$ supersymmetry in nine spacetime
dimensions with Lorentz-invariant central charges. These central
charges are encoded in a  
two-by-two real symmetric  matrix $Z^{ij}$, which can be decomposed as
\be
Z^{ij} = b\,\d^{ij} + a \,(\cos\theta \,\sigma_3 + \sin \theta\,
\sigma_1)^{ij} \,.
\ee
Here $\sigma_{1,3}$ are the real symmetric Pauli matrices. 
We note that the central charge associated with the parameter $a$
transforms as a doublet under the SO(2) group that rotates the two 
supercharge spinors, while the central charge proportional to 
the parameter $b$ is SO(2)
invariant. Subsequently one shows that BPS states that carry these 
charges must satisfy the mass formula,
\be 
M_{\rm BPS} = \vert a \vert + \vert b\vert \,.
\ee
Here one can distinguish three types of BPS supermultiplets. One type 
has central charges $b=0$ and $a\not=0$. These are 
$1/2$-BPS muliplets, because they are annihilated by half of the
supercharges. The KKA supermultiplets that comprise KK states of IIA
supergravity are of this type.  Another type of $1/2$-BPS mulitplets
has central charges $a=0$ and $b\not =0$. The KKB supermultiplets
that comprise the KK states of IIB supergravity are of this
type. Finally there are $1/4$-BPS multiplets 
(annihilated by one fourth of the supercharges) 
characterized by the fact that neither $a$ nor $b$ vanishes. 

For type-II string theory one obtains these central charges in
terms of the left- and right-moving momenta, $p_{\rm L}$, $p_{\rm R}$,
that characterize winding and momentum along $S^1$. However, the
result takes a different form for the IIA and the IIB theory, 
\be
Z^{ij} = \left\{\begin{array}{lr}
\ft12(p_{\rm L}+p_{\rm R}) \d^{ij} +  \ft12(p_{\rm L}-p_{\rm R})
\sigma_3^{ij} \,,\quad & {\rm (for \;IIB)}\\[3mm]
 \ft12(p_{\rm L}-p_{\rm R}) \d^{ij} +  \ft12(p_{\rm L}+p_{\rm R})
\sigma_3^{ij} \,.\quad & {\rm (for \;IIA)}
\end{array} \right.
\ee
The corresponding BPS mass formula is equal to
\be
M_{\rm BPS} = \ft12\vert p_{\rm L}+p_{\rm R}\vert  +  \ft12\vert
p_{\rm L}-p_{\rm R}\vert \,.
\ee
For $p_{\rm L}=p_{\rm R}$ we confirm the original identification of
the momentum states, namely that IIA momentum states constitute KKA
supermultiplets, while IIB momentum states constitute KKB
supermultiplets. For the winding states, where $p_{\rm L}=-p_{\rm R}$,
one obtains the opposite result: IIA winding states constitute KKB
supermultiplets, while IIB winding states constitute KKA 
supermultiplets. The $1/4$-BPS multiplets arise for string states that have
either right- or left-moving oscillator states, so that either $M_{\rm
BPS} = \vert p_{\rm L}\vert$ or $\vert p_{\rm R}\vert$ with $p_{\rm L}^{\,2}
\not= p_{\rm R}^{\,2}$. All of this is entirely consistent with 
T-duality\cite{DHS,DLP},
according to which there exists a IIA and a IIB perspective, with 
decompactification radii are that inversely proportional and with an
interchange of winding and momentum states.

It is also possible to view the central charges from the perspective of the
eleven-dimensional (super)membrane. Assuming that the two-brane charge
takes values in the compact coordinates labeled 
by 9 and 10, which can be generated by wrapping the membrane
over the corresponding $T^2$, one readily finds the expression,
\be 
Z^{ij} = Z_{9\,10} \,\d^{ij} - (P_9 \,\sigma_3^{ij} - P_{10} \,
\sigma_1^{ij})\,.
\ee
When compactifying on a torus with modular parameter
$\tau$ and area $A$, the BPS mass formula takes the form  
\bea
M_{\rm BPS} &=& \sqrt{P_9^{\,2} + P_{10}^{\,2} } + \vert
Z_{9\,10}\vert \nonumber\\
&=& {1\over \sqrt{A\,\tau_2}} \vert q_1 + \tau \,q_2\vert + T_{\rm m}
A \,\vert p\vert \,. \label{BPS-membrane}
\eea
Here $q_{1,2}$ denote the momentum numbers on the torus and $p$ is the
number of times the membrane is wrapped over the torus; $T_{\rm m}$
denotes the supermembrane tension. 
Clearly the KKA states correspond to the momentum modes on $T^2$ while
the KKB states are associated with the wrapped membranes on the
torus. Therefore there is a rather natural way to describe the IIA and IIB
momentum and winding states starting from a (super)membrane in eleven
spacetime dimensions. This point was first emphasized in ref.~15.
% \cite{JHS}. 
%%%%%%%%%%%%%%%%%%%%%%%%%%%%%%%%%%%%%%%%%%%%%%%%%%%%%%%%%%%%%
 
%%%%%%%%%%%%%%%%%%%%%%%%%%%%%%%%%%%%%%%%%%%%%%%%%%%%%%%%%%%%%
\section{BPS-extended Supergravity}
\noindent
%%%%%%%%%%%%%%%%%%%%%%%%%%%%%%%%%%%%%%%%%%%%%%%%%%%%%%%%%%%%%
It is possible to consider $N=2$ supergravity in nine spacetime
dimensions and couple it to the simplest BPS supermultiplets
corresponding to the KKA and KKB states. As shown in the previous
section there are three central charges and nine-dimensional
supergravity possesses precisely three gauge fields that couple to these
charges. From the perspective of eleven-dimensional supergravity
compactified on $T^2$, the  KK states transform as KKA
multiplets. Their charges transform obviously with respect to an SO(2)
associated with 
rotations of the coordinates labeled by 9 and 10. Hence we have a
double tower of these charges with corresponding 
KKA supermultiplets. On the other hand, from the perspective of IIB 
compactified on $S^1$, the KK states constitute KKB multiplets
and their charge is SO(2) invariant. Here we have a single tower of KKB
supermultiplets. However, from the perspective of nine-dimensional
supergravity one is led to couple both towers of KKA and KKB
supermultiplets simultaneously. In that case one obtains some
dichotomic theory\cite{ADLN}, which we refer to as BPS-extended 
supergravity. In the
case at hand this new theory describes the 
ten-dimensional IIA and IIB theories in certain decompactification
limits, as well as eleven-dimensional supergravity. But the theory is
in some sense truly twelve-dimensional with three compact coordinates,
although there is no twelve-dimensional Lorentz invariance, not even
in a uniform decompactification limit, as the fields never depend on
all the twelve coordinates! Whether this kind of BPS-extended
supergravity offers a viable scheme in a more general context than the
one we discuss here, is not 
known. In fact, not much work has been done on incorporating the
coupling of BPS multiplets into a supersymmetric field
theory. Nevertheless, in the case at 
hand we know a lot about these couplings from our knowledge of
the $T^2$ compactification of eleven-dimensional supergravity and the
$S^1$ compactification of IIB supergravity. 

The fields of nine-dimensional $N=2$ supergravity are listed in
table~1, where we also indicate their relation with the fields of
eleven-dimensional and ten-dimensional IIA/B supergravity upon
dimensional reduction. It is not necessary to work out all the 
nonlinear field redefinitions here, as the corresponding fields can be
uniquely identified by their scaling weights under SO(1,1), a symmetry
of the massless theory that emerges upon dimensional reduction and  is 
associated with scalings of the internal vielbeine. The scalar field
$\sigma$ is related to $G_{99}$, the IIB metric component in the
compactified dimension, by $G_{99}=\exp(\sigma)$; likewise it
is related to the 
determinant of the eleven-dimensional metric in the compactified
dimensions, which is equal to $\exp(-\ft43\sigma)$. The precise relationship
follows from  comparing the SO(1,1) weights through the
dimensional reduction of IIB and eleven-dimensional supergravity. In
nine dimensions supergravity has two more scalars, 
which are described by a nonlinear sigma model based on 
${\rm SL}(2,{\bf R})/{\rm SO}(2)$. The coset is described by the
complex doublet of fields $\phi^\a$, which satisfy a constraint
$\phi^\a\phi_\a =1$ and are subject to a local SO(2) invariance, so
that they describe precisely two scalar degrees of freedom
($\a=1,2$). We expect that the local SO(2) invariance 
can be incorporated in the full BPS-extended supergravity theory and
can be exploited in the construction of the couplings of the various
BPS supermultiplets to supergravity. 

We already mentioned  the three abelian vector gauge
fields which couple to the central charges. There are two vector
fields $A_\m^\a$, which are the KK 
photons from the $T^2$ reduction of eleven-dimensional supergravity
and which couple therefore to the KKA states. From the IIA perspective
these correspond to the KK states on $S^1$ and the D0
states. From the IIB side they 
originate from the tensor fields, which confirms that they couple to
the IIB (elementary and D1) winding states. These two fields transform 
under SL(2), which can be understood {from} the perspective of the
modular transformation on $T^2$ as well as from the S-duality
transformations that rotate the elementary strings with the D1 strings.
The third gauge field, denoted by $B_\m$, is a singlet under SL(2)
and is the KK photon on the IIB side, so that it couples
to the KKB states. On the IIA side it originates from the IIA tensor
field, which is consistent with the fact that the IIA winding states
constitute KKB supermultiplets. 

{From} the perspective of the supermembrane, the KKA states are the
momentum states on $T^2$, while the KKB states correspond to the
membranes wrapped around the torus. While it is gratifying to see how
all these correspondences work out, we stress that from the
perspective of nine-dimensional $N=2$ supergravity, coupled to the
smallest $1/2$-BPS states, the results follow entirely from
supersymmetry.  

%%%%%%%%%%%%%%%%%%%%%%%%%%%%%%%%%%%%%%%%%%%%%%%%%%%%%%%%%%%%%%%%%%
\begin{table}
\tcaption{The bosonic fields of the eleven dimensional, type-IIA,  
nine-dimensional $N=2$ and type-IIB  supergravity theories. 
The eleven-dimensional and ten-dimensional indices, respectively, are 
split as $\hat{M} =(\mu, 9,10)$ and $M=(\mu,9)$, where $\mu = 0,1,\ldots 8$. 
The last column lists the SO(1,1) scaling weights of the fields.}
\vspace{2mm}
\begin{center}
\begin{tabular}{ccccr}
\hline $D=11$ & IIA & $D=9$ & {~}\quad IIB \quad{~} & SO(1,1) \\ 
\hline \hline \\[-3mm]%%%%%%%%%%%%%%%%%%%%%%%%%%%%%%%%%%%%%%%%%%%%%%%%%%%%%
$\hG_{\mu \nu}$ & $G_{\m \n}$  & $g_{\m \n}$  & $G_{\m \n}$ & $0\;\,$ \\[1mm]
\hline \\[-3mm]%%%%%%%%%%%%%%%%%%%%%%%%%%%%%%%%%%%%%%%%%%%%%%%%%%%%%%%%%%%%
$\hat{A}_{\mu {\scriptscriptstyle \,9\,10}}$  & $C_{\mu
{\scriptscriptstyle\,9}}$ &  $B_{\m}$ & $G_{\m {\scriptscriptstyle\,9}}$
& $-4\;\,$\\[1mm] 
\hline \\[-3mm]%%%%%%%%%%%%%%%%%%%%%%%%%%%%%%%%%%%%%%%%%%%%%%%%%%%%%%%%%%%%%
$\hG_{\mu {\scriptscriptstyle\,9}}$, $\hG_{\mu
{\scriptscriptstyle\,10}}$  & $G_{\mu {\scriptscriptstyle\,9}}$ ,
$C_{\mu}$  & $A_{\mu}^{\, \alpha}$ & $A_{\m
{\scriptscriptstyle\,9}}^{\, \alpha}$ & $3\;\,$  \\[1mm]  
\hline \\[-3mm]%%%%%%%%%%%%%%%%%%%%%%%%%%%%%%%%%%%%%%%%%%%%%%%%%%%%%%%%%%%%
$\hat{A}_{\m \n {\scriptscriptstyle\,9}}$, $\hat{A}_{\m \n
{\scriptscriptstyle\,10}}$ & $C_{\m \n {\scriptscriptstyle\,9}}, C_{\m
\n}$ &  $A_{\m\n}^{\,\a}$ &  $A_{\m\n}^{\,\a}$  & $-1\;\,$  \\ [1mm]
\hline \\[-3mm]%%%%%%%%%%%%%%%%%%%%%%%%%%%%%%%%%%%%%%%%%%%%%%%%%%%%%%%%%%%%
$\hA_{\m \n\rho}$  & $C_{\m \n \rho}$  & $A_{\m\n\rho}$  &
$A_{\m\n\rho\sigma}$ & $2\;\,$  \\[1mm] 
\hline \\[-3mm]%%%%%%%%%%%%%%%%%%%%%%%%%%%%%%%%%%%%%%%%%%%%%%%%%%%%%%%%%%%%%
$\hat G_{{\scriptscriptstyle 9 \,10}}$, $\hat{G}_{{\scriptscriptstyle9\,
9}}$, $\hat G_{{\scriptscriptstyle 10\,  10}}$ & $\phi$,  
$G_{{\scriptscriptstyle 9\,9}}$, $C_{\scriptscriptstyle  9}$ &
$\left\{ \begin{array}{l} \phi^\a  \\[1mm] \exp (\sigma ) \end{array}
\right. $ & $\begin{array}{l} \phi^\a \\[1mm] G_{{\scriptscriptstyle
9\,9}} 
\end{array}$ & $\begin{array}{r} $0$ \\[1mm]  $7$ \end{array}$
\\[1mm]\\[-3mm]  
\hline %%%%%%%%%%%%%%%%%%%%%%%%%%%%%%%%%%%%%%%%%%%%%%%%%%%%%%%%%%%%
\end{tabular}
%\vspace{.3cm}
%\caption{d type-IIB  supergravity theories. 
\end{center}

\end{table}
%%%%%%%%%%%%%%%%%%%%%%%%%%%%%%%%%%%%%%%%%%%%%%%%%%%%%%%%%%%%%%%%%%%

The resulting BPS-extended theory incorporates eleven-dimensional
supergravity and the two type-II supergravities in special
decompactification limits. But, as we stressed above, we are dealing
with a twelve-dimensional theory here, of which three coordinates are
compact, except that no field can depend on all of the three compact
coordinates. The theory has obviously two mass scales associated with
the KKA and KKB states. We return to them in a moment. Both S- and
T-duality are manifest, although the latter has become trivial as
the theory is not based on a specific IIA or IIB perspective. One
simply has the freedom to view the theory from a IIA or a IIB
perspective and interpret it accordingly. 

We should discuss the fate of the group ${\rm G}= {\rm SO}(1,1) \times
{\rm SL}(2,{\bf R})$ of pure supergravity after coupling the theory to
the BPS multiplets. The central charges of the BPS states form a
discrete lattice, which is affected by this group. Hence, after
coupling to the BPS states, we only have a discrete subgroup that
leaves the charge lattice invariant. This is the group ${\rm
SL}(2,{\bf Z})$.  

The KKA and KKB states and their interactions with the massless theory
can be understood from the perspective of compactified  eleven-dimensional and IIB
supergravity. In this way we are able to deduce the following BPS mass
formula, 
\be
M_{\rm BPS}(q_1,q_2,p) = m_{\scriptscriptstyle\rm KKA} \,{\rm
e}^{3\sigma/7} \,\vert q_\a \phi^\a \vert +  m_{\scriptscriptstyle\rm
KKB} \,{\rm e}^{-4\sigma/7} \,\vert p\vert\,,
\ee
where $q_\a$ and $p$ refer to the integer-valued KKA and KKB charges,
respectively, and $m_{\scriptscriptstyle\rm KKA}$ and
$m_{\scriptscriptstyle\rm KKB}$ are two independent  mass scales. 
This formula can be compared to the membrane BPS formula
\eqn{BPS-membrane} in the eleven-dimensional frame. One then finds
that 
\be
m^2_{\scriptscriptstyle \rm KKA}\, m^{~}_{\scriptscriptstyle\rm KKB}
\propto  T_{\rm m}\,,
\ee
with a numerical proportionality constant. 

%%%%%%%%%%%%%%%%%%%%%%%%%%%%%%%%%%%%%%%%%%%%%%%%%%%%%%%%%%%%%
\section{Supertraces and $R^4$-terms}
\noindent
%%%%%%%%%%%%%%%%%%%%%%%%%%%%%%%%%%%%%%%%%%%%%%%%%%%%%%%%%%%%%
In this BPS-extended supergravity theory one can integrate out the BPS 
supermultiplets and study their contribution to the effective 
action. In this way one makes contact with the $R^4$-terms that were
considered by refs.~19-21.% \cite{green,green1,green2}. 
This amounts to
a one-loop calculation in a nine-dimensional field theory. Since we
evaluate the coefficient 
of the $R^4$ term at zero momentum, the relevant amplitude has the
structure of a box diagram in massive $\varphi^3$ theory in nine spacetime
dimensions (relying on the usual recombination with the non-box
graphs that often appears in gauge theories). The  contributions from
the KKA and KKB states are equal to\cite{DWL} 
\be
A_4^{{\scriptscriptstyle\rm KKA} + {\scriptscriptstyle\rm KKB}} = 
{1\over (2\pi)^9}\int\;{\rm d}^9k\;\sum_{q_1,q_2,p}\; {1 \over
[k^2+M^2(q_1,q_2,p)]^4} \,.
\ee
The momentum integral is linearly divergent and it is multiplied by a
so-called helicity supertrace $B_n$, with $n=8$, which correctly
counts all the virtual 
states belonging to the KKA and KKB supermultiplets. These helicity
supertraces can be calculated straightforwardly and depend on the
supermultiplet of virtual states in the loop. One finds that 
$1/2$-BPS multiplets have a zero supertrace when $n<8$, the
$1/4$-BPS multiplets have a zero supertrace when  $n<12$, and the
generic supermultiplets have a zero supertrace 
for $n<16$. Therefore only the KKA and KKB supermultiplets are expected to
contribute to the $R^4$-terms (at one loop). 

After a Poisson resummation this result can be written as follows
(we use the BPS mass formula \eqn{BPS-membrane}), 
\bea
A_4^{\scriptscriptstyle{\rm KKA} + {\scriptscriptstyle\rm KKB}} &=& 
{2\over 3}\,{A^{-1/2} \over (4\pi)^6 } \;
\sum_{q_1^\prime,q_2^\prime}\,{\tau_2^{3/2}\over  \vert
q_1^\prime+\tau q_2^\prime \vert^3} 
+  {4\over 3}\,{T_{\rm m}\, A  \over
(4\pi)^6 } \; \sum_{p^\prime}\,{1\over  p^{\prime \,2}}
\nonumber\\ 
&=& {2\over 3}\,{1 \over
(4\pi)^6 } \; \Big[ A^{-1/2} \, f(\tau,\bar \tau) + \ft23
\pi^2\,T_{\rm m}\,A \Big] \,. 
\eea
In the last line we dropped the terms in the
sum for  $q_1^\prime=q_2^\prime=0$ and $p^\prime=0$. These terms
represent 
the ultraviolet divergences. The modular function $f(\tau,\bar\tau)$
is defined by\cite{green}  
\be 
f(\tau,\bar \tau) =  \sum_{(q_1^\prime,q_2^\prime)\neq
(0,0)}\,{\tau_2^{3/2}\over  \vert q_1^\prime+\tau q_2^\prime \vert^3}
\,.
\ee
This result is invariant (as it should) under the IIB S-duality symmetry 
$\tau\rightarrow (a\tau +b)/(c\tau+d)$ with $a$, $b$, $c$ and $d$
integers satisfying $ac-bd=1$. The contribution from the KKB
states is such that the result is compatible with T-duality of type-II
string theory.

It is a gratifying result that including the KKA and the KKB states in
the context of a uniform regularization scheme leads directly to a
result that is consistent with T-duality. Nevertheless, there remains
a number of critical questions. The degree of 
divergence  of the initial box 
graphs is not reduced by supersymmetry, because the relevant helicity
supertraces for the KKA and the KKB multiplets, which represent the sum
over the virtual states, yield identical 
multiplicative factors. The subtraction dependence also reflects
itself in the fact that the KKA contributions disappear in the
decompactification limit to eleven dimensions, $A\to\infty$, which is
counterintuitive.  Hence the result remains in principle subtraction 
dependent. In fact the uniform regularization implies that the sum
over the KKA and the KKB states leads to double counting of the
massless states. 

%%%%%%%%%%%%%%%%%%%%%%%%%%%%%%%%%%%%%%%%%%%%%%%%%%%%%%%%%%%%%
\section{Concluding remarks}
\noindent
%%%%%%%%%%%%%%%%%%%%%%%%%%%%%%%%%%%%%%%%%%%%%%%%%%%%%%%%%%%%%
We have considered extensions of supergravity where one couples to a
supersymmetric set of BPS states that is complete with respect to the
U-duality group. We used maximal  nine-dimensional supergravity as an
example, coupled to both KKA and KKB supermultiplets. From the point of view
of U-duality, this example is not quite satisfactory in demonstrating
all relevant features, as the KKA and the KKB states transform
separately under U-duality. In lower-dimensional spacetimes the
situation is different and the corresponding BPS states transform
irreducibly under U-duality. Nevertheless, in this way one obtains
a dichotomic theory that can be regarded as a twelve-dimensional field
theory, which, upon suitable decompactification limits, leads to 
eleven-dimensional supergravity or ten-dimensional IIA/B
supergravity. 

There are interesting questions regarding the field-theoretic coupling
of the fields associated with BPS states. In the case at hand
these can in principle be answered, because the couplings can be deduced
from the coupling of the massive KK fields in the compactifications of
eleven-dimensional and IIB supergravity. One such questions concerns
the role of the local symmetry ${\rm H}={\rm SO}(2)$ that one uses in
the description of the SL(2)/SO(2) coset space for the nonlinear sigma
model. This symmetry has a (composite) gauge field, which does not
describe additional degrees of freedom. 

What is difficult to answer at present is the question whether such
BPS-extended theories can be fully
consistent as local field theories. It is known that the coupling of
fields of higher spins can lead to minimal-coupling
inconsistencies. As long as we consider terms that are at most 
quadratic in the massive fields, there seems no
immediate discrepancy. However, once one includes the higher-order
couplings between the BPS states, this is not so clear. Yet, it would
be interesting to calculate these couplings as in this way  one can 
make contact with the algebra of BPS states discussed in ref.~23.
% \cite{hm}. 

As we mentioned before, the extension with BPS states is 
interesting from the perspective 
of alternative formulations of eleven-dimensional supergravity which
have been written down long ago\cite{deWitNicolai,Nicolai}. There it
was found that the eleven-dimensional theory does exhibit nontrivial
traces of the hidden symmetry groups ${\rm E}_{7(7)}$ and ${\rm
E}_{8(8)}$, although the invariance under these groups is only
realized upon truncating to the massless states. However, to some
extent the 
lack of invariance can be traced back to the incompleteness of the BPS
states with respect to U-duality (for a recent discussion, see
ref.~11).
% \cite{istanbul}). 
Therefore it seems reasonable
to expect that there do exist interacting field theories of
supergravity coupled to BPS states which are consistent and respect
U-duality and therefore are suitable starting points for learning more
about M-theory.  On the other hand, we do expect certain obstacles to
this program for the case of toroidal compactifications to low
spacetime dimensions. First of all, we already 
know of some inconsistencies in the U-duality assignments of the
central charges, 
and secondly, the BPS charges will be mutually nonlocal. For instance,
the pointlike charges in $D=4$ spacetime dimensions correspond to
electric and magnetic charges, which can not be incorporated in the
framework of a local field theory.     

%%%%%%%%%%%%%%%%%%%%%%%%%%%%%%%%%%%%%%%%%%%%%%%%%%%%%%%%%%%%%
\nonumsection{Acknowledgements}
\noindent
I thank Ivan Herger, Dieter L\"ust, Hermann Nicolai and Niels Obers 
for valuable discussions. 

%%%%%%%%%%%%%%%%%%%%%%%%%%%%%%%%%%%%%%%%%%%%%%%%%%%%%%%%%%%%%%%
%%%%%%%%%%%%%%%%%%%%%%%%%%%%%%%%%%%%%%%%%%%%%%%%%%%%%%%%%%%%%%%

%%%%%%%%%%%%%%%%%%%%%%%%%%%%%%%%%%%%%%%%%%%%%
%%%%%%%%%%%%%%%%%%%%%%%%%%%%%%%%%%%%%%%%%%%%%

\begin{thebibliography}{000}
%
\bibitem{BST}
E. Bergshoeff, E. Sezgin and P.K. Townsend,
%{\em Supermembranes and Eleven-Dimensional Supergravity},
{\bibit Phys.\ Lett.}\ {\bf B189} (1987) 75.       
%
\bibitem{DWHN}
B. de Wit, J. Hoppe and H. Nicolai,
%{\em On the Quantum Mechanics of Supermembranes},
{\bibit Nucl.\ Phys.}\ {\bf B305} (1988) 545.
%
\bibitem{BFSS}
T. Banks, W. Fischler, S.H. Shenker and L. Susskind,
%{\em M-Theory as a Matrix Model: a Conjecture},
{\bibit Phys.\ Rev.}\ {\bf D55} (1997) 5112, hep-th/9610043.
%
\bibitem{corfu}
H. Nicolai and R. Helling, {\bibit Supermembranes and M(atrix)
Theory}, in proc. of the Trieste String School on non-perturbative
aspects of 
String Theory and Supersymmetric Gauge Theories, 1998, hep-th/9809103;\\
B. de Wit, {\bibit Supermembranes and Super Matrix Models}, in
{\bibit Quantum Aspects of Gauge Theories, Supersymmetry and  
Unification},  proc. Curfu, 1998, eds. A. Ceresole, C. Kounnas,
D. L\"ust and S. Theisen (Lecture notes in Physics 525, Springer,
1999), hep-th/9902051;\\
W. Taylor, {\bibit The M(atrix) Model of M Theory}, lectures given at
the NATO Advanced Study Institute on Quantum Geometry, Iceland, 1999,
hep-th/0002016. 
%
\bibitem{HullTownsend} C.M. Hull and P.K. Townsend, {\bibit
Nucl. Phys.} {\bf B438} (1995) 109, hep-th/9410167. 
%
\bibitem{CremmerJulia} E. Cremmer and B. Julia, {\bibit Phys. Lett.}
{\bf 76B} (1978) 409.  
%
\bibitem{ObersPioline} N.A. Obers and B. Pioline, {\bibit
Phys. Rept.} {\bf 318} (1999) 113, {hep-th/9809039} 
%
\bibitem{CJS} E. Cremmer, B. Julia and J. Scherk, 
%{\it Supergravity Theory in Eleven Dimensions}, 
{\bibit Phys. Lett.} {\bf 76B} (1978) 409. 
%
\bibitem{Townsend} 
P.K. Townsend, {\bibit Phys. Lett.} {\bf B350} (1995) 184,
hep-th/9501068.  
\bibitem{Witten}
E. Witten,
%{\em String Theory Dynamics in Various Dimensions},
{\bibit Nucl. Phys.} {\bf B443} (1995) 85, hep-th/9503124.
%
\bibitem{istanbul} B. de Wit and H. Nicolai, {\bibit Hidden
Symmetries, Central Charges and All That}, to be published in the
proceedings of the G\"ursey Memorial Conference II, June 2000, Istanbul.
%
\bibitem{deWitNicolai} B. de Wit and H. Nicolai, {\bibit Phys. Lett.}
{\bf 155B} (1985) 47; {\bibit Nucl. Phys.} {\bf B274} (1986) 363.
%
\bibitem{Nicolai} H. Nicolai,  {\bibit Phys. Lett.} {\bf 187B} (1987) 363.
%
\bibitem{BHO} E. Bergshoeff, C. Hull and T. Ortin, {\bibit
Nucl. Phys.} {\bf  B451} (1995) 547, hep-th/9504081. 
%
\bibitem{JHS} J.\ H.\ Schwarz, 
{\bibit An $SL(2,Z)$ Multiplet of Type-II Superstrings}, 
{\bibit Phys.\ Lett.}\ {\bf B360} (1995) 13, 
Erratum: {\em ibid}, {\bf B364} (1995) 252, hep-th/9508143, 
{\bibit Nucl.\ Phys.\ Proc.\ Suppl.}\ {\bf 49} (1996) 183, hep-th/9509148.
%
\bibitem{ADLN}
M. Abou-Zeid, B. de Wit, D. L\"ust and H. Nicolai,
%{\em Space-Time Supersymmetry, IIA/B Duality and M-Theory},
{\bibit Phys. Lett.} {\bf 466B} (1999) 144, hep-th/9908169.
%
\bibitem{DHS} M.\ Dine, P.\ Huet and N.\ Seiberg, 
%{\em Large and Small Radius in String Theory}, 
{\bibit Nucl.\ Phys.}\ {\bf B322} (1989) 301.                 
%
\bibitem{DLP} J.\ Dai, R.\ G.\ Leigh and J.\ Polchinski, 
%{\em New Connections Between String Theories}, 
{\bibit Mod.\ Phys.\ Lett.}\ {\bf A4} (1989) 2073. 
% 
\bibitem{green} M.B. Green, M. Gutperle and P. Vanhove,
%{\em One Loop in Eleven Dimensions}, 
{\bibit Phys.\ Lett.}\ {\bf B409} (1997) 177,
hep-th/9706175.
%
\bibitem{green1}
M.B. Green and M. Gutperle,
%{\em Effects of D-Intantons},
{\bibit Nucl. Phys.} {\bf B498} (1997) 195,
hep-th/9701093.
%
\bibitem{green2}
M.B. Green and P. Vanhove,
%{\em D-Instantons, Strings and M-Theory},
{\bibit Phys.\ Lett.}\ {\bf B408} (1997) 122,
hep-th/9704145;
%{\em The Low Energy Expansion of the One-Loop
%Type II Superstring Amplitude},
{\bibit Phys. Rev.} {\bf D61} (2000) 104011, hep-th/9910056.
%
\bibitem{DWL} B. de Wit and D. L\"ust, {\bibit Phys. Lett.} {\bf B477}
(2000) 299, hep-th/9912225.
%
\bibitem{hm}
J. Harvey and G. Moore,
%{\em Algebras, BPS States and Strings},
{\bibit Nucl. Phys.} {\bf B463} (1996) 315, hep-th/9510182.

\end{thebibliography}
\end{document}